# High-resolution borehole earthquake monitoring at San Andreas Fault Observatory at Depth, Parkfield, California

*Ruiqing He and Björn Paulsson*

*Paulsson, Inc., 16543 Arminta Street, Van Nuys, California 91406, USA*

**Abstract**

Downhole earthquake monitoring, without the complex effects from the near surface, can record more and better seismic data than monitoring on surface. The San Andreas Fault Observatory at Depth (SAFOD) is a borehole observatory equipped with different instruments inside to study the earthquake mechanism of the San Andreas fault at Parkfield, California. During April to May in 2005, Paulsson deployed an 80-level 3-component geophone array in the SAFOD main hole, and continuously recorded seismic data for about 13 days. We located 125 local earthquakes from the borehole earthquake monitoring data using a homogeneous velocity model and compared it with 35 earthquakes' locations from surface earthquake monitoring by the United State Geological Survey (USGS) during the same monitoring time. The borehole earthquake locating is assumably more accurate in the borehole's vicinity. We also compared the result with 1,074 earthquakes' locations from the surface earthquake monitoring in the last 9 years from 2015 to 2024. The hypocenters from our nearly 2 weeks' borehole earthquake monitoring form similar structures as that from the 9 years' surface earthquake monitoring by the USGS.

**Introduction**

The near surface in most areas on the Earth is a complex barrier for recording earthquakes because of its weathering layers. Its effects include attenuation and bending of the seismic waves that the earthquakes generate. Although expensive, downhole earthquake monitoring is necessary in some areas where earthquakes need to be closely monitored.

San Andreas Fault (SAF) is one of the most active and destructive faults in the world. It stretches about 1,200 km through California from southeast to northwest. Parkfield is in the middle of the SAF, and near several M6 earthquakes' epicenters, and frequently undergoes M2 earthquakes. This makes Parkfield an ideal place to study the earthquake mechanism of the SAF. The San Andreas Fault Observatory at Depth (SAFOD) is a borehole observatory deployed with different instruments inside to study earthquake-related phenomenon. During April to May in 2005, Paulsson installed a 4,000 ft long 80-level 3-component (3C) geophone array in the SAFOD main hole (a 3 km deviated well), and continuously recorded seismic data for about 13 days (with a few breaks). Besides this, Duke university had installed a 32-level 3C geophone array in the adjacent (sharing the same well pad) SAFOD pilot hole (a 2 km vertical well) and monitored earthquakes for several years (Oye and Ellsworth, 2005).

The Parkfield earthquake experiment started in the 1980s, and by then the SAF was the most densely monitored (and by the best equipment) fault on the Earth (Harris and Arrowsmith, 2006). The SAFOD project was initiated in the late 1990s and has provided scientists with insightful information about the SAF's rock properties, geological structures, and seismic activities (Hickman, 2007). All the seismic data we used for this study, from both surface and borehole earthquake monitoring can be downloaded from https://ds.iris.edu/.

The Paulsson recorded borehole seismic data at the SAFOD has been used by Chavarria et al. (2007) for fault zone characterization, by Rentsch et al. (2010) to locate about 2 dozen micro earthquake events, and by Reshetnikov et al. (2010) to image the SAF. Despite the high Signal/Noise (S/N) ratio, the borehole seismic data has not been thoroughly exploited, especially since there are few publications demonstrating the differences between surface and borehole earthquake monitoring at Parkfield. In this study, we located 125 local earthquakes using the borehole earthquake monitoring data and compared the result with 35 earthquakes' locations during the same 13 days' period in 2005 and 1,074 earthquakes' locations in the last 9 years from 2015 to 2024 using surface earthquake monitoring by the United State Geological Survey (USGS).

## Downhole Geophones' Orientation through A PAME

Figure 1 shows the locations of the SAFOD site and offset check shots at 13 different locations. In the 3D model, the sparse dots represent the previous settings of the borehole geophone array, and the dense dots represent the final positions of the 80-level 3C geophones. The deepest geophone was at 9,000 ft measured depth. Large charges of 80 lb. Pentolite along 100 ft depth near the surface were detonated as the offset check shots.

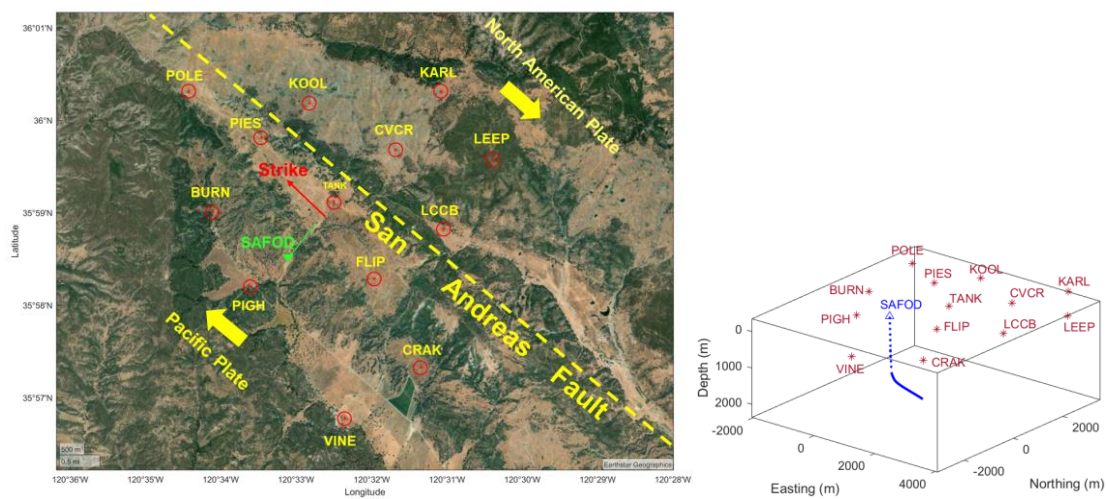

Fig. 1: The map (left) of the 13 offset check shots around the SAFOD site, and a 3D illustration (right) of the check shots (red stars) and the SAFOD main hole, displayed with equal scales in the easting, northing, and vertical directions.

The main purpose of the check shots was to calibrate the downhole radial-component geophones' orientation. Instead, here we used a ParAxial Micro Earthquake (PAME) event that we recorded to orient the downhole geophones. A PAME is a micro earthquake at a close distance to the geophones and near the extended path of the borehole's trajectory. It is characterized by strong direct P wave and weak direct S wave in the axial-component geophones, and weak direct P wave but strong direct S wave in the radial-component geophones. Figure 2 shows the PAME we used to orient the downhole radial-component geophones, and Figure 3 shows its waveforms after the orientation which maximizing the direct S wave onto one of the radial components which is the Strike direction nearly parallel to the strike direction of the SAF. The Planar direction here is defined as the downward direction orthogonal to both the borehole axial direction and the Strike direction. The benefit of using the PAME for the downhole geophones' orientation is that its distance to the bottom geophone is 1/5 of the bottom geophone's distance to its nearest check shot on the surface.

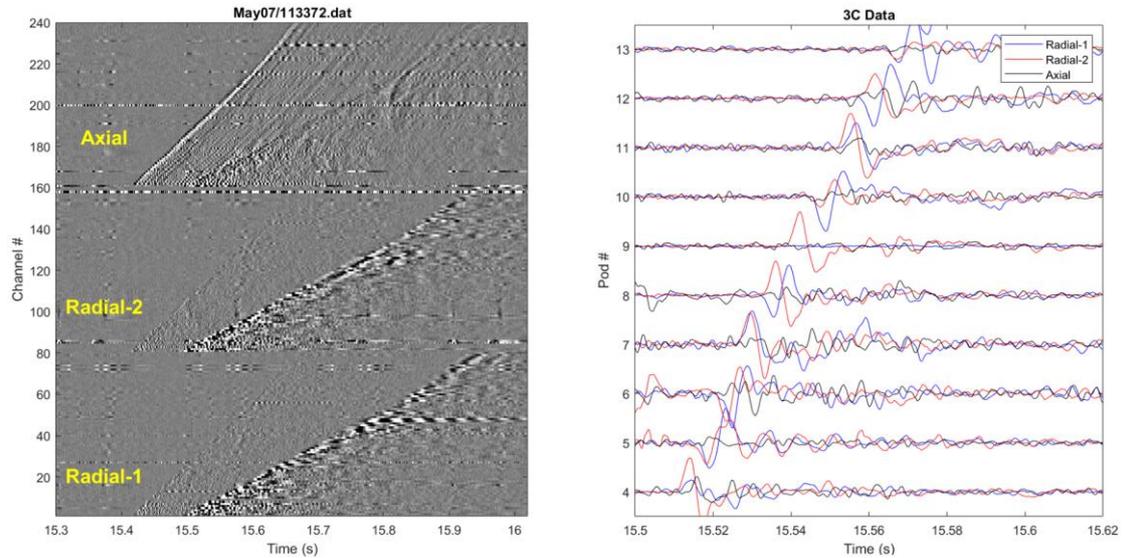

Fig. 2: The 3C seismic data of a PAME (left) and some of its selected direct S-wave waveforms (right).

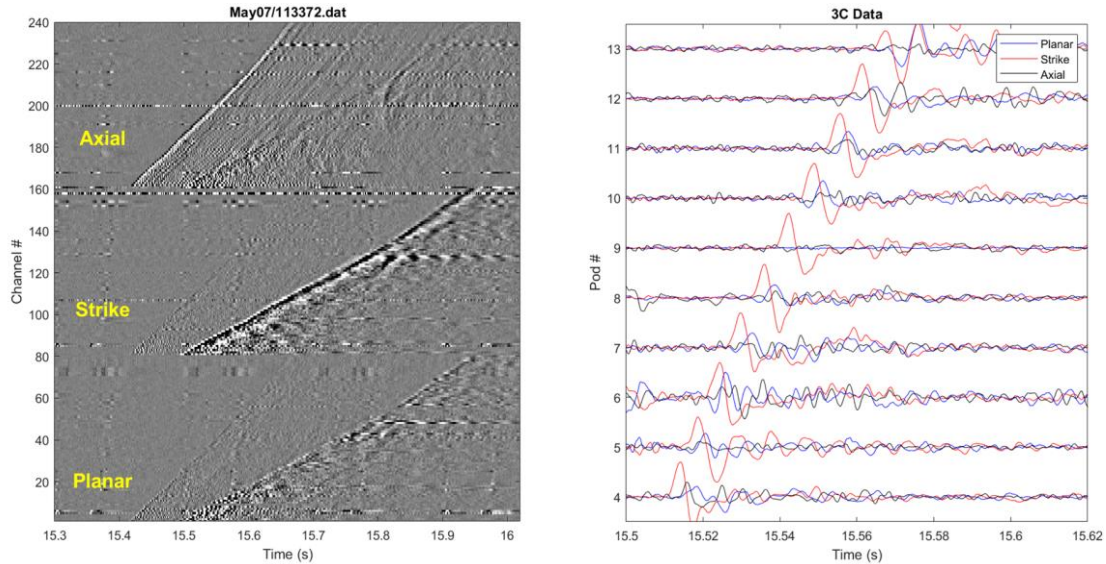

Fig. 3: The oriented seismic data (left) and some of its selected direct S-wave waveforms (right) of the PAME shown in figure 6.

### Check Shots Being Checked for Borehole Earthquake Locating

After the downhole radial-component geophones have been oriented through a PAME, we tested the capability of borehole earthquake locating using the offset check shots' borehole seismic data. The earthquake locating algorithm is first to find the overall direct P wave's direction from the 3C borehole seismic data and locate the earthquake's source location in a homogeneous model with P-wave velocity of 4,650 m/s and S-wave velocity of 2,620 m/s.

Figure 4 shows the comparison of the located check shots' source positions vs. the actual positions. Some have small while others have considerable errors in the lateral directions, and most have considerable errors in the vertical direction. After investigations, we concluded that the reasons for such errors are the following in the order of severity: 1) strong vertical velocity variations near the surface; 2) lack of clear direct S waves in the seismic data; 3) refraction of the direct P waves; 4) lack of a 3D velocity model. Natural earthquakes usually are distant from the surface and characterized by strong S waves, which makes borehole earthquake locating assumably much more reliable than in this test.

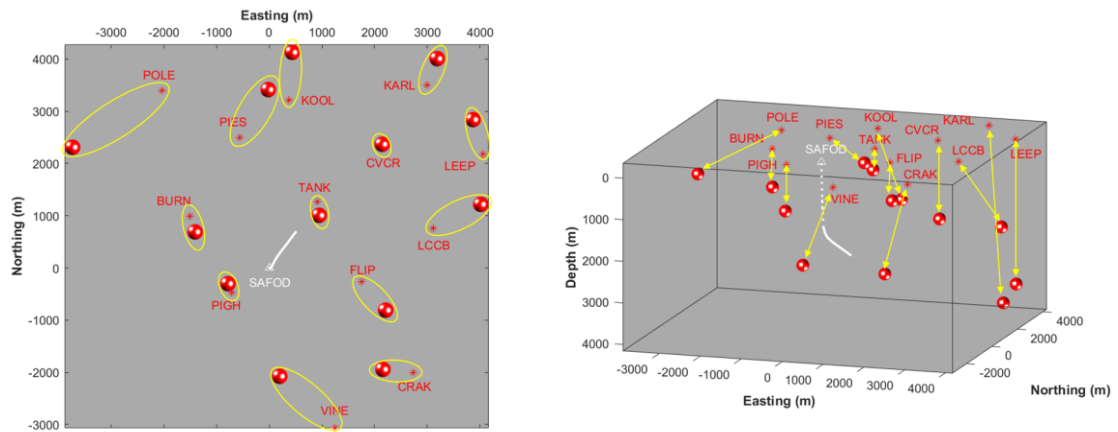

Fig. 4: The top view (left) and a 3D view (right) of the located check shots' source positions (red balls) by using the borehole seismic data with a homogeneous velocity model. The locating errors are mostly because of strong vertical velocity variations near the surface and lack of clear direct S waves in the explosive check shots' seismic data.

Figure 5 shows the results of a reciprocal earthquake experiment in which the actual geophones are assumed to be virtual earthquake sources, and the actual check shot sources are assumed to be virtual geophones. From the direct P waves' first arrival time contours, the epicenters of the virtual earthquakes would have the minimal direct P wave arrival times if a homogeneous velocity model is assumed. From the figure, the assumedly located epicenters (big pink stars) deviate from the virtual epicenters (small red stars) by considerable distances.

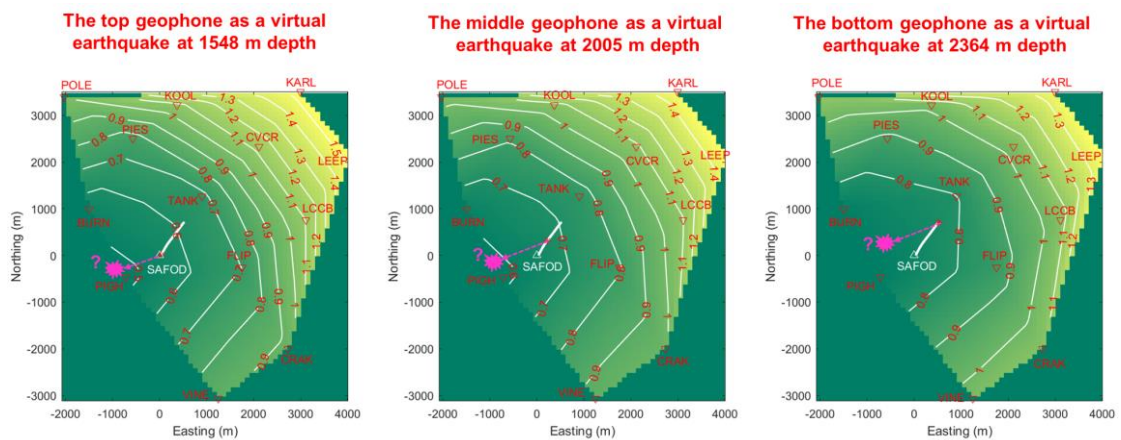

Fig. 5: A reciprocal earthquake experiment with the actual geophones being virtual earthquake sources and the actual check shots being virtual geophones. From the direct P waves' first arrival time contours, with a homogeneous velocity model, the pink stars would represent the located epicenters.

## Hypocenters by Borehole Earthquake Locating at SAFOD

Figure 6 shows the 125 hypocenters located by Paulsson borehole earthquake monitoring for about 13 days in 2005 with the reference of the offset check shots' locations. It shows most of the hypocenters are on the SAF's fault plane which is nearly vertical with a tiny tilt angle toward the southwestern direction.

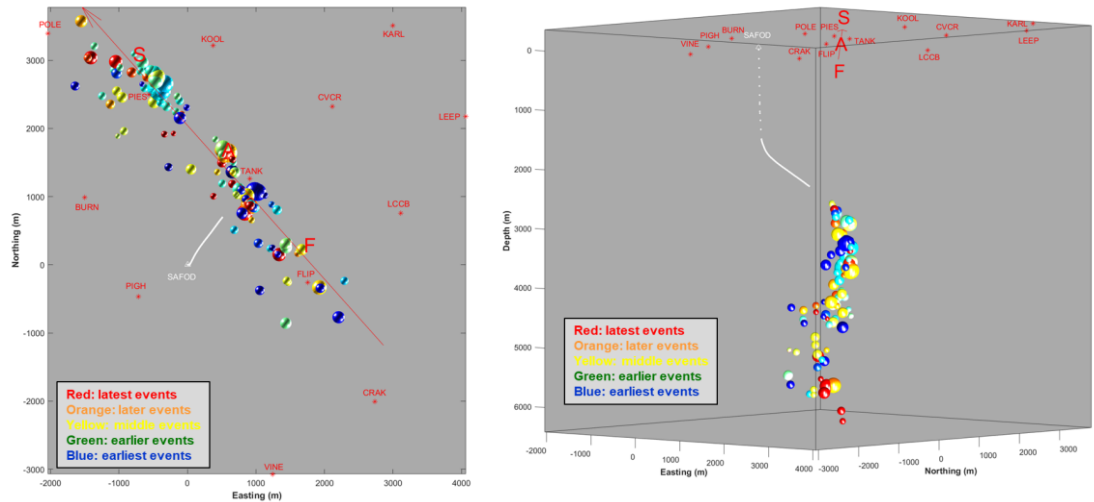

Fig. 6: The top view (left) and a 3D view (right) of the Paulsson located hypocenters of the recorded 125 earthquakes by borehole earthquake monitoring in about 13 days in 2005. The colors of the balls represent the timing of the earthquakes, and the sizes of the balls represent the magnitudes of the earthquakes ranging from M(-3) to M3.

As a comparison, Figure 7 shows the 35 hypocenters located by the USGS using surface earthquake monitoring at the same time as Paulsson borehole earthquake monitoring in 2005.

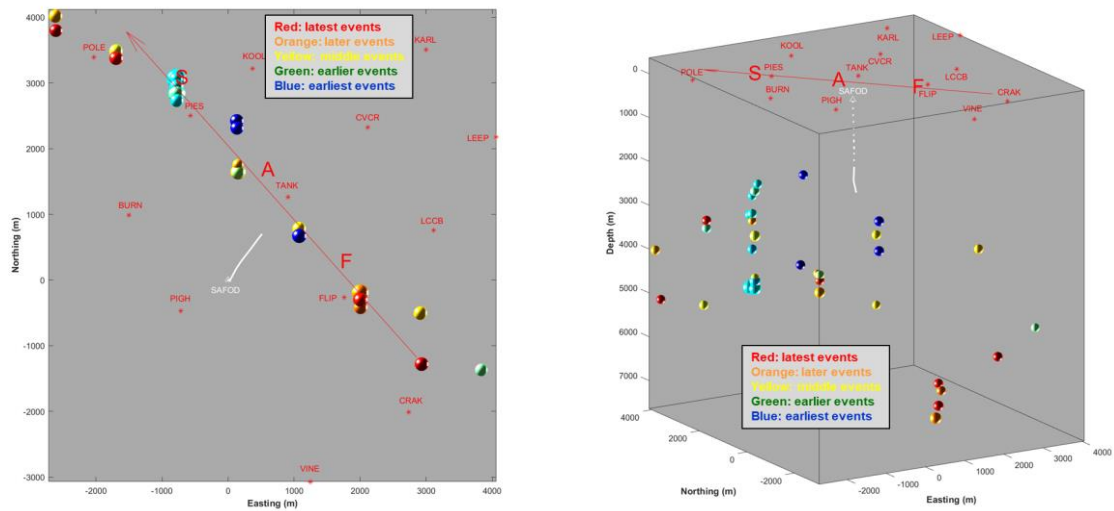

Fig. 7: The top view (left) and a 3D view (right) of the USGS located hypocenters of the recorded 35 earthquakes by surface earthquake monitoring at the same time as the borehole earthquake monitoring in about 13 days in 2005. The colors of the balls represent the timing of the earthquakes, and the sizes of the balls represent the magnitudes of the earthquakes ranging from M0 to M3.

Figure 8 shows the hypocenter comparison of an earthquake located by both the borehole and surface earthquake monitoring. Both epicenters are close to each other and not far from the borehole, but the depths are 1.2 km different. The borehole seismic data of this earthquake shows clear direct P and S waves, which makes the hypocenter located by the borehole earthquake monitoring assumably more reliable.

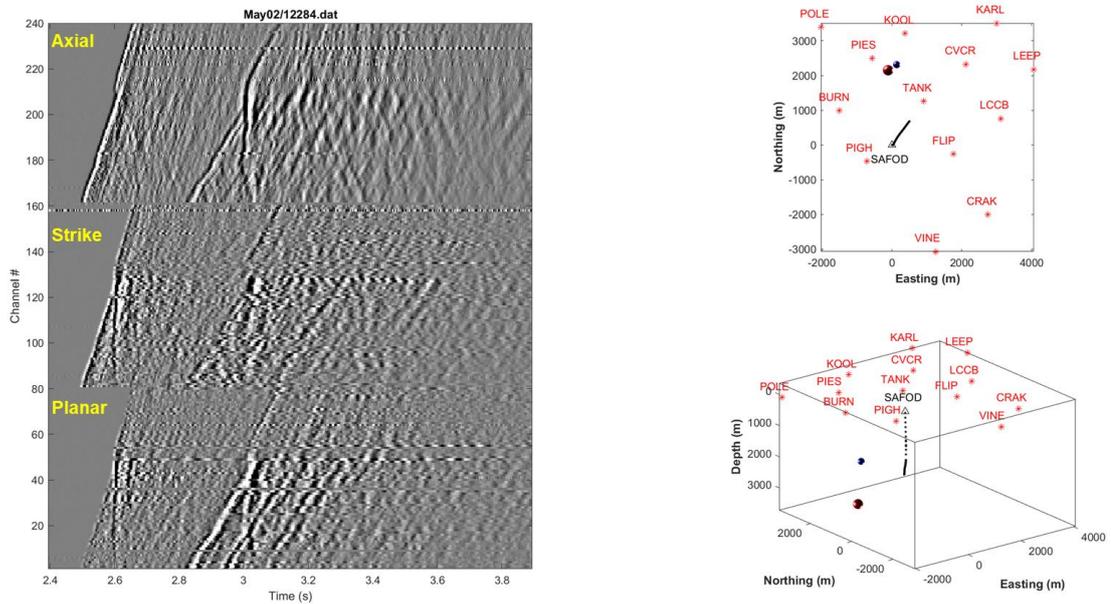

Fig. 8: The top view (right top) and a 3D view (right bottom) of the hypocenter comparison of an earthquake between Paulsson's borehole earthquake monitoring (the red ball) and the USGS's surface earthquake monitoring (the blue ball). The left panel shows the borehole seismic data of the earthquake which shows clear direct P wave and S wave. The hypocenter located by the borehole earthquake monitoring is assumably more reliable in this case.

Figure 8 shows the comparison of the 1,074 hypocenters located by the USGS in the last 9 years from October 2015 to October 2024 using surface earthquake monitoring vs. that by Paulsson in about 13 days in 2005 using borehole earthquake monitoring. Both results show similar structures of the hypocenters' distribution.

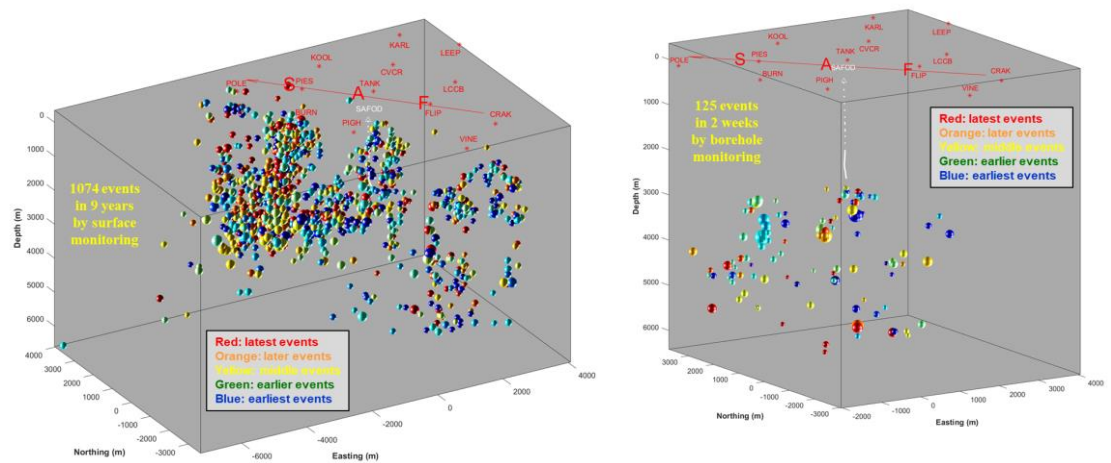

Fig. 9: The comparison of hypocenters located by the USGS in the last 9 years from 2015 to 2024 using surface earthquake monitoring (left) vs. that by Paulsson in about 13 days in 2005 using borehole earthquake monitoring (right). Both results show similar structures of the hypocenters' distribution.

## Discussions

In this study, using a homogeneous velocity model, we conveniently located the local earthquakes recorded by a borehole earthquake monitoring in the SAFOD main hole for about 13 days in 2005. The result is assumably more reliable than the surface earthquake monitoring by the USGS during the same time. The located 125 hypocenters from our borehole earthquake monitoring in less than 2 weeks also take a similar shape as the 1,074 located hypocenters from the USGS's surface earthquake monitoring in the last 9 years from 2015 to 2024.

The USGS's surface earthquake locating algorithm assumedly has used a 3D velocity model, otherwise it could have considerable errors as demonstrated by our reciprocal earthquake experiment using the offset check shots' borehole seismic data. An accurate subsurface 3D velocity model is not only difficult to obtain, but also difficult to verify, especially in complex areas like at the SAFOD site. Borehole earthquake locating, because of the shorter distances to the local earthquakes, is less affected by the inaccuracy of the subsurface 3D velocity model than surface earthquake locating. Before a verified 3D velocity model is available, we should bear in mind that the far offset earthquakes in the borehole earthquake monitoring may be affected by the refracted direct P waves. An optimal solution to this would be monitoring earthquakes along the fault with multiple boreholes.

Besides that, borehole earthquake monitoring records more micro earthquakes than surface earthquake monitoring. In this study, it is estimated that the borehole earthquake monitoring has recorded at least 4 times as many micro earthquakes as surface earthquake monitoring. This number does not include other many smaller seismic events that the borehole earthquake monitoring has recorded but not used in this locating study.

As we know, most major earthquakes at one place repeat with intermissions over several decades or even centuries. This is longer than most people's years of working experience or even lifespans. In order to accelerate earthquake research and in greater details, borehole earthquake monitoring is a proven solution.

## Acknowledgements

We thank NSF and other sponsors for supporting the Parkfield earthquake experiments, especially the SAFOD project.

## Author Contributions

Ruiqing He contributed all the data processing and most of the writing of the manuscript.

Björn Paulsson designed the borehole seismic array, the experiment, and data acquisition, and contributed part of the writing of the manuscript.